\documentclass[aps,prd,preprint,superscriptaddress,amsmath,amssymb,showpacs]{revtex4-1}
\usepackage{dcolumn}
\usepackage{graphicx}
\usepackage{float}
\usepackage{physics}
\usepackage[colorlinks=true,allcolors=blue]{hyperref}

\begin{document}

\title{Thermodynamics of heavy quarkonium in magnetic field background  }

\author{Jing Zhou}
\email{171001005@stu.njnu.edu.cn}
\affiliation{Department of Physics, Nanjing Normal University, Nanjing, Jiangsu 210097, China}
\author{Xun Chen}
\email{chenxunhep@qq.com}
\affiliation{Key Laboratory of Quark and Lepton Physics (MOE) and Institute of Particle Physics,\\
Central China Normal University, Wuhan 430079, China}
\author{Yan-Qing Zhao}
\email{yanqzhao@qq.com}
\affiliation{Key Laboratory of Quark and Lepton Physics (MOE) and Institute of Particle Physics,\\
Central China Normal University, Wuhan 430079, China}
\author{Jialun Ping}
\email{jlping@njnu.edu.cn}
\affiliation{Department of Physics, Nanjing Normal University, Nanjing, Jiangsu 210097, China}

%\date{\today}

\begin{abstract}
We study the effect of magnetic field on heavy quark-antiquark pair in both Einstein-Maxwell(EM) and Einstein-Maxwell-Dilaton(EMD) model. The interquark distance, free energy, entropy, binding energy and internal energy of the heavy quarkonium are calculated. It is found that the free energy suppresses and the entropy increases quickly with the increase of the magnetic field $B$. The binding energy vanishes at smaller distance when increasing the magnetic field, which indicates the quark-antiquark pair dissociates at smaller distance. The internal energy which consists of free energy and entropy will increase at large separating distance for non-vanishing magnetic field. These conclusions are consistent both in the EM and EMD model. Moreover, we also find that the quarkonium will dissociate easier in the parallel direction than that in the transverse direction for EMD model, but the conclusion is opposite in EM model. Lattice results are in favor of EMD model. Besides, a Coulomb-plus-linear potential(Cornell potential) can be realized only in EMD model. Thus, a dilaton field is proved to be important in holographic model. Finally, we also show that the free energy, entropy and internal energy of a single quark in EMD model with the presence of magnetic field.
\end{abstract}


\maketitle

\section{Introduction}\label{sec:01_intro}
It is believed that the so called quark gluon plasma(QGP) has been created in relativistic heavy ion collisions at the Relativistic Heavy Ion Collider (RHIC) and the Large Hadron Collider (LHC)\cite{Adcox:2004mh,Adams:2005dq,Arsene:2004fa,Miller:2007ri}. Studying the property of the QGP help us to have further understanding of the nature. Heavy quarkonium, which is a bound state of a quark and its antiquark, are among the most sensitive probes used in the experimental study of the QGP and its properties. The information encoded in heavy quarkonia observables supplements penetrating electromagnetic probes and hard (jet) probes and the rich flow observables, thus complementing each other in characterizing the dynamics of quarks and gluons up to the final hadronic states. Since heavy quarks emerge essentially in early hard processes, they witness the course of a heavy-ion collision -- either as individual entities or subjects of dissociating and regenerating bound states.\cite{Zollner:2020nnt} Depending on the temperature of QGP or size of the quarkonium state, the heavy quark and antiquark may be screened from each other which affects the production rates of heavy quarkonia in heavy-ion collisions\cite{Matsui:1986dk}.

The QGP created in heavy-ion collisions has indeed been found to be strongly coupled at the experimentally accessible temperatures somewhat above the critical temperature\cite{Adcox:2004mh,Adams:2005dq,Arsene:2004fa,Miller:2007ri}, then the non-perturbative way become valid. The AdS/CFT correspondence\cite{Maldacena:1997re,Witten:1998qj,Gubser:1998bc} or the gauge/gravity duality can provide important information to study the strongly coupled systems. The unique advantage of holography leads us to study the various properties of the QGP with gravitational methods. And this holography has given many important insights to study the different nature of strongly coupled matter.

As we know that the strong magnetic fields play essential roles in the noncentral heavy ion collisions\cite{Kharzeev:2007jp,Kharzeev:2004ey,Fukushima:2008xe,Gatto:2012sp,Abelev:2009ac,Abelev:2009ad,Adamczyk:2013hsi,Adamczyk:2013kcb,Hirano:2010je,Abelev:2012pa,Mo:2013qya,Zhong:2014cda,Zhong:2014sua,Feng:2016srp}. For instance, the magnetic fields produced at the top collision energies of the Large Hadron Collider is the order of $eB \sim 15 m_{\pi}^2 \sim 0.3GeV^2$. Such strong fields may have consequences for the transport and thermodynamic properties of the QGP formed in later stages of heavy ion collisions. As the knowledge of the stability of heavy quarkonia has important implications on the fate of heavy quarkonia in the QGP, it is important to study the stability of bound meson states in this magnetic field range. In holographic model, the presence of magnetic field is introduced by adding a $U(1)$ gauge field which is dual of the electromagnetic current in the gravitational action. For our current purposes, we will just introduce a constant magnetic field $B$, namely, we have no interest in the fluctuations of the Abelian gauge field. Note that this $B$ is the 5-dimensional magnetic field which needs to be suitably rescaled via the AdS length $L_{AdS}$ to get the physical, 4-dimensional, magnetic field $B$. How to do this can be found in \cite{Dudal:2015wfn}. To approach a real QCD, a dilaton field(a running coupling constant) is often introduced to break the conformal symmetry. In early year, this dilaton field is introduced by hand in an ad hoc way into the metric. However, this dilaton is self-consistently solved from the gravity equations in recent years such as EMD model. In this paper, we both calculate the EM and EMD model as a comparison and clarify the necessity of the dilaton field. The QCD phase transition in the presence of magnetic field has been studied in Ref.\cite{Mamo:2015dea,Rougemont:2015oea,Li:2016gfn,Rodrigues:2017iqi,Rodrigues:2018pep,He:2020fdi}. The holographic entanglement entropy with a magnetic field has been investigated in\cite{Dudal:2016joz,Fujita:2020qvp}. The heavy quark diffusion with a magnetic field has been reported in \cite{Dudal:2018rki}. The energy loss of heavy and light quarks in holographic magnetized background has been discussed in\cite{Zhu:2019ujc}. And other relative works can be found in\cite{Braga:2018zlu,DElia:2018xwo,Fukushima:2012kc,Ferreira:2014kpa,Bali:2011qj,Evans:2016jzo,
Ballon-Bayona:2017dvv,Gursoy:2016ofp,Gursoy:2017wzz}.

Moreover, many works use AdS/CFT correspondence to investigate the free energy or potential of a heavy quark-antiquark pair in strongly interacting matter\cite{Andreev:2006eh,Colangelo:2010pe,Li:2011hp,Herzog:2006ra}. In fact, the binding energy can be regarded as the energy difference of free energy of quark-antiquark pair and two free quarks. The difference between free energy and binding energy has been discussed in\cite{Ewerz:2016zsx}. Besides the free energy, the entropy would be responsible for melting the quarkonium and may be related to the nature of confinement and deconfinement. Lattice QCD studies show that the additional entropy associated with the presence of static quark-antiquark pair in the QCD plasma\cite{Kaczmarek:2002mc,Kaczmarek:2004gv}. It is found that the entropy will grow with the inter-quark distance, and give rise to the entropic forces that tend to destroy the quark-antiquark pair\cite{Kharzeev:2014pha,Satz:2015jsa}. Entropic destruction results in an anomalously strong quarkonium suppression in the temperature range near $T_c$(deconfinement temperature). The heavy quark entropy in strong magnetic fields from holographic black hole engineering has been studied in \cite{Critelli:2016cvq}. Thermal entropy of a quark-antiquark pair above and below deconfinement has been discussed in \cite{Dudal:2017max}. More interesting works can be seen in\cite{Petreczky:2005bd,Fadafan:2015ynz,Chen:2017lsf,Zhang:2016fwr,Zhang:2017izi,Zhang:2020upv,Arefeva:2018hyo,Arefeva:2018cli}.

The rest of the paper is organized as follows: In Sec.~\ref{sec:02}, we will give a brief introduction on the Einstein-Maxwell system and thermodynamic relationship with the magnetic field. In Sec.~\ref{sec:03}, we calculate the results of quark-antiquark distance, free energy, entropy, binding energy and internal energy with magnetic field background and discuss the effect of magnetic field in both transverse and parallel to magnetic field direction from the Einstein-Maxwell model. Similarly, we can compute all these quantities in the Einstein-Maxwell-Dilaton system in Sec.~\ref{sec:04}. The free energy, entropy and internal energy of a single quark with the presence of magnetic field have been computed in Sec.~\ref{sec:05}. Finally, the conclusion and discussion will be given in Sec.~\ref{sec:06}.

\section{The Setup}\label{sec:02}

The action of the gravity background with back-reaction of magnetic field in the
Einstein-Maxwell (EM) system \cite{Mamo:2015dea,Dudal:2015wfn,Li:2016gfn} is given as:

\begin{equation}
S= \int \frac{1}{16\pi G_{5}}\sqrt{-g}(R-F^{MN}F_{MN}+\frac{12}{L_{AdS}^{2}}) d^{5}x.
\end{equation}
where $R$ is the scalar curvature, $G_{5}$ is 5D Newton constant, $g$ is the determinant of metric $g_{\mu\nu}$, $L_{AdS}$ is the AdS radius and $F_{MN}$ is the tensor of the U(1) gauge field.
The Einstein equation for the EM system could be derived as follows,
\begin{equation}
E_{MN}-\frac{6}{L_{AdS}^{2}}g_{MN}-2(g^{IJ}F_{MJ}F_{NJ}-\frac{1}{4}F_{IJ}F^{IJ}g_{MN})=0.
\end{equation}
where $E_{MN}=R_{MN}-\frac{1}{2} R g_{MN}$. $E_{MN}$, $R_{MN}$ are Einstein tensor and the Ricci tensor. We take $L_{AdS} =1$ in the following sections. The ansatz of metric is taken as,
\begin{equation}
ds^{2}=\frac{1}{z^{2}}(-f(z)dt^{2}+\frac{1}{f(z)}dz^{2}+h(z)(dx^{2}_{1}+dx^{2}_{2})+q(z)dx^{2}_{3}).
\end{equation}
A constant magnetic field is along the $x_3$ direction in this metric. For a black hole solution, we have $f(z = z_{h}) = 0$ at horizon $z = z_{h}$. $q(z)$ and $h(z)$ are regular function of z in the region $0 < z < z_{h}$.  As discussed in Ref.\cite{Li:2016gfn}, we will only take the leading expansion as
\begin{equation}
\begin{aligned}
f(z)&=1-\frac{z^{4}}{z^{4}_{h}}(1-\frac{2}{3}B^{2}z^{4}_{h}\log(\frac{z}{z_{h}})),\\
q(z)&=1+\frac{2}{3}B^{2}\log(z)z^{4},\\
h(z)&=1-\frac{1}{3}B^{2}z^{4}\log(z).
\end{aligned}
\end{equation}
Note that the arguments of the logs actually are dimensionless. $z_h$ is divided by $L_{AdS}$ which is set to one in this paper. The Hawking temperature and the magnetic field $B$ is given as
\begin{equation}
T=\frac{1}{\pi z_{h}}-\frac{B^2 z^{3}_{h}}{6 \pi}.
\end{equation}
one can take proper values of $z_{h}$ and $B$ to set the temperature $T$ and magnetic field $B$ in the dual 4D theory. Actually, it is found that the approximate of leading order Eq.(4) is good enough for $T\geq0.15 GeV$ and $B\leq0.15 \mathrm{GeV}^2$ \cite{Li:2016gfn,Zhu:2019ujc,Feng:2019boe,Zhu:2019igg}. For $T \sim 0.15$ GeV and $B \sim 0.15$ GeV$^2$, one finds $z_h \sim 2$ GeV$^{-1}$ and therefore we anticipate subleading terms in the expansion will be suppressed by a power of $B^4 z_h^8 \sim \frac{1}{8}$.

The Nambu-Goto action of the worldsheet in the Minkowski metric is given by
\begin{equation}
    S_{NG} = - \frac{1}{2\pi\alpha'} \int d^{2}\xi \sqrt{- \det g_{ab}},
\end{equation}
where $g_{ab}$ is the induced metric, $\frac{1}{2\pi\alpha'}$ is the string tension and
\begin{equation}
    g_{ab} = g_{MN} \partial_a X^M \partial_b X^N, \quad a,\,b=0,\,1.
\end{equation}
Here, $X^M$ and $g_{MN}$ are the coordinates and the metric of the AdS space.

Then, the Nambu-Goto action can be rewritten as
\begin{equation}
    S_{NG} = - \frac{R^2}{2\pi\alpha'T}\int_{-L/2}^{L/2} d{x}^{2} \sqrt{g_1(z) \frac{d{z}^2}{d{x}^2} + g_2(z)}.
\end{equation}
So, we can parameterize in parallel direction with $\xi^0 = t,\,\xi^1 = x_3$. $g^{par}_1(z)$, $g^{par}_2(z)$ are
\begin{gather}
    g^{par}_1(z) =\frac{1}{z^{4}}, \\
    g^{par}_2(z) =\frac{1}{z^{4}}f(z)q(z).
\end{gather}
And for transverse direction, we set $\xi^0 = t,\,\xi^1 = x_1$. Thus, $g^{tra}_1(z)$, $g^{tra}_2(z)$ are

\begin{equation}
\begin{aligned}
    g^{tra}_1(z)&=\frac{1}{z^{4}},\\
    g^{tra}_2(z)&=\frac{1}{z^{4}}f(z)h(z).
\end{aligned}
\end{equation}

The separating distance of $Q\bar{Q}$ pair is
\begin{equation}
    L = 2\int_0^{L/2} dx = 2\int_0^{z_0} \frac{dx}{dz} dz =2 \int_0^{z_0} [\frac{g_2(z)}{g_1(z)} (\frac{g_2(z)}{g_2(z_0)}-1)]^{-1/2} d{z}.
\end{equation}
And the renormalized free energy of $Q\bar{Q}$ pair can be written as

\begin{equation}
    \frac{\pi F_{Q\bar{Q}}}{\sqrt{\lambda}} = \int_0^{z_0} d{z} (\sqrt{\frac{g_2(z)g_1(z)}{g_2(z)-g_2(z_0)}}-\sqrt{g_2(z\rightarrow 0)}) - \int_{z_0}^\infty \sqrt{g_2(z\rightarrow 0)} d{z}\label{13}.
\end{equation}

The entropy of $Q\bar{Q}$ pair is given as

\begin{equation}
    S_{Q\bar{Q}} = - \frac{\partial F_{Q\bar{Q}}}{\partial T} = -\frac{\partial F_{Q\bar{Q}}} {\partial z_h}\frac{\partial z_h}{ \partial T},
\end{equation}

where $T$ is the temperature of the QGP. And the binding energy is defined as $E_{Q \bar{Q}} = F_{Q \bar{Q}}-2F_{Q }$\cite{Ewerz:2016zsx}, equivalently,

\begin{equation}
    \frac{\pi E_{Q\bar{Q}}}{\sqrt{\lambda}} = \int_0^{z_0} d{z} (\sqrt{\frac{g_2(z)g_1(z)}{g_2(z)-g_2(z_0)}} - \sqrt{g_2(z\rightarrow 0)}) - \int_{z_0}^{z_h} \sqrt{g_2(z\rightarrow 0)} d{z}.
\end{equation}

 The binding energy $E_{Q\bar{Q}}$ = 0, when the free energy of the interacting quark pair equals the free energy of non-interacting heavy quarks pair. The free energy of a single quark can be calculated by (\ref{13}), but let upper limit of integral is $z_h$. Namely,

\begin{equation}
    \frac{F_{Q}}{\sqrt{\lambda}} = \frac{1}{2\pi}(\int_0^{z_{h}} d{z} (\sqrt{g_1(z)}-\frac{1}{z^{2}}) - \frac {1}{z_{h}}).
\end{equation}

 We set the constant $\sqrt{\lambda} =1$ for convenient in this paper. According to the Ref.\cite{Bali:2014kia}, the internal energy of the quark-antiquark pair can be calculated by $U_{Q \bar{Q}}=F_{Q \bar{Q}}+T S_{Q \bar{Q}}$. If we consider the total energy of the system, we should add $M B$ term, where $M$ represents the magnetization which is associated to $B$. But we only consider the energy of quark-antiquark pair in our paper.

\section{THERMODYNAMIC QUANTITIES OF HEAVY QUARKONIUM In Einstein-Maxwell (EM) system WITH MAGNETIC FIELD}\label{sec:03}
\begin{figure}[H]
    \centering
    \includegraphics[width=16cm]{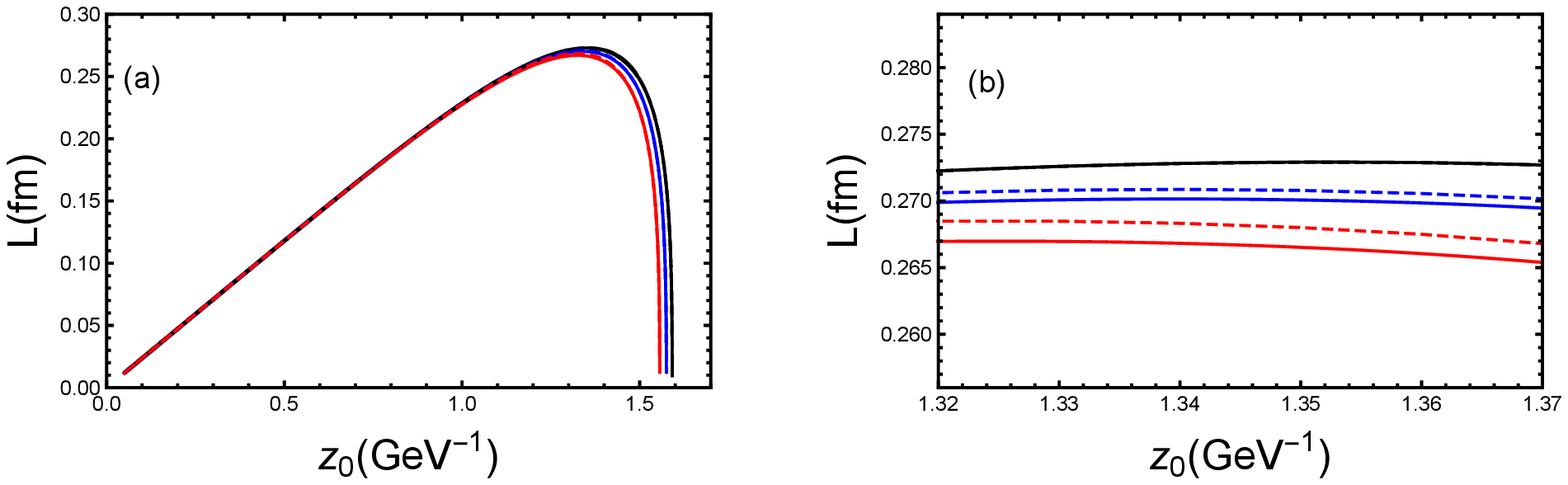}
    \caption{\label{L12}(a) The dependence of interquark distance $L$ of $Q\bar{Q}$  pair on $z_0$ with different magnetic fields. Black line is $B = 0 \mathrm{GeV}^2$, blue line is $B = 0.1 \mathrm{GeV}^2$ and red line is $B = 0.15 \mathrm{GeV}^2$ for vanishing chemical potential. The solid line is for transverse direction and dashed line is for parallel direction.(b) is a partly enlarged view of (a).}
\end{figure}

Fig.~\ref{L12} shows the dependence of interquark distance $L$ of $Q\bar{Q}$  pair on $z_0$ for different $B$. One can find that the interquark distance $L$ increases with $z_{0}$. After $L$ reaches the maximum value, then it goes down to zero. It indicates that the $Q\bar{Q}$ pair is in the deconfinement phase. This is because that there is a maximum of the separating distance $L_S$ in the deconfinement phase, which is the screening distance $L_S$. If we continue to increase the $z_{0}$, $U$-shape strings become unstable, then it will go down to zero. The unstable branch is not physically favored, we only focus our discussion on $L \leq L_S$ in this paper. Moreover, we can see that larger magnetic field will lead to easier dissolution of the heavy quarkonium. And the heavy quarkonium dissociate easier in the transverse magnetic direction than that in the parallel magnetic direction.

\begin{figure}[H]
    \centering
    \includegraphics[width=16cm]{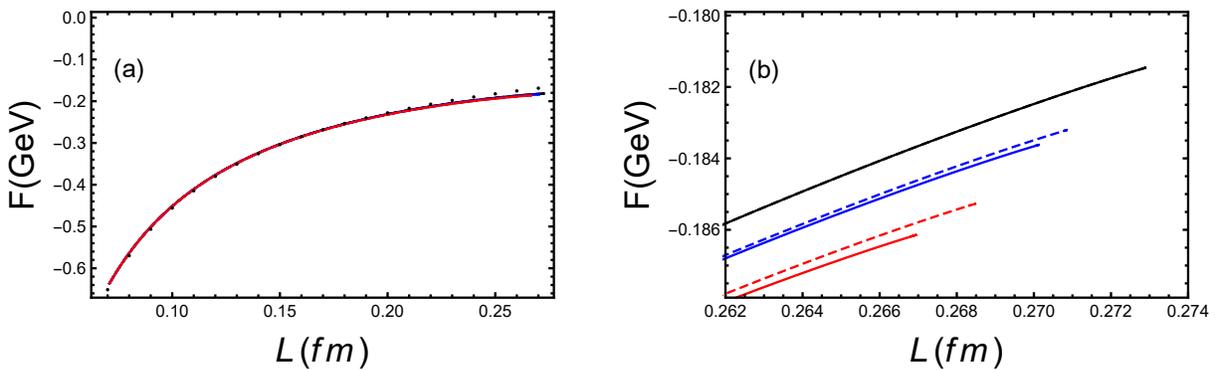}
    \caption{\label{F12}The free energy of quarkonium on $L$ at different magnetic fields. Black line is $B = 0 \mathrm{GeV}^2$, blue line is $B = 0.1 \mathrm{GeV}^2$ and red line is $B = 0.15 \mathrm{GeV}^2$ for vanishing chemical potential. The solid line is for transverse direction and dashed line is for parallel direction. The dot is a fit of the free energy with Coulomb potential. (b) is a partly enlarged view of (a).}
\end{figure}

The dependence of free energy of heavy $Q\bar{Q}$ pair on the interquark distance $L$ for different $B$ is plotted in Fig.~\ref{F12}. It is shown that the free energy is only a Coulomb potential which lacks a linear potential at large $L$. When adding the dilaton, however, free energy will have a linear potential at large $L$ which will be presented in the next section. To be more clear, we also fit the free energy with $F = - \frac{0.0456}{L}$ in Fig.~\ref{F12}. We can see that only a Coulomb potential is enough to describe the behavior. And one can find that the free energy will be suppressed with the increase of magnetic field. Further, the magnetic field in transverse direction will suppress the free energy more than in parallel direction. But lattice results\cite{Bonati:2016kxj} show magnetic field of parallel direction will suppress the free energy more than the transverse direction. These two questions will be solved by adding a dilaton field in the next section.

\begin{figure}[H]
    \centering
    \includegraphics[width=16cm]{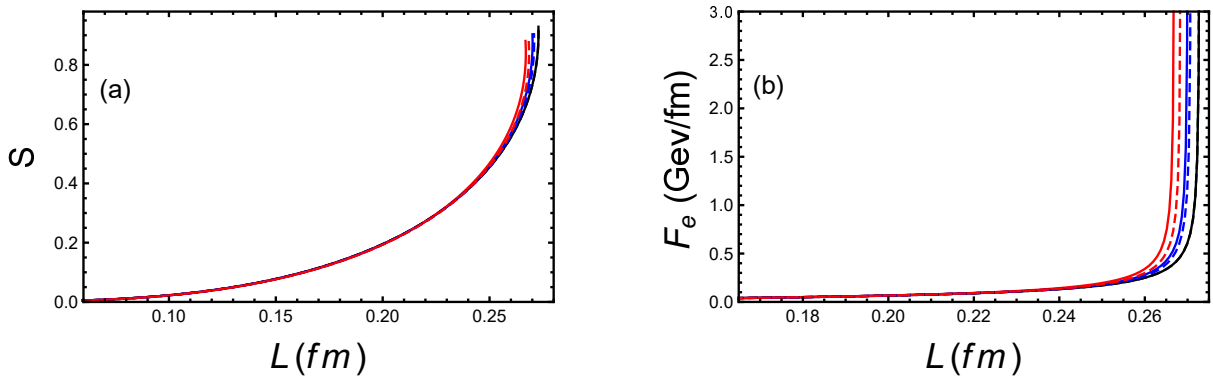}
    \caption{\label{S12} The entropy of quarkonium on $L$ at different magnetic fields. Black line is $B = 0 \mathrm{GeV}^2$, blue line is $B = 0.1 \mathrm{GeV}^2$ and red line is $B = 0.15 \mathrm{GeV}^2$ for vanishing chemical potential. The chemical potential is $0$ and the temperature is $T = 0.2\,\mathrm{GeV}$. (b) Corresponding entropic force as a function of L. The solid line is for transverse direction and dashed line is for parallel direction.
    }
\end{figure}

Then, we show the dependence of entropy of heavy $Q\bar{Q}$ pair on the interquark distance $L$ at different magnetic fields $B$ in Fig.~\ref{S12}(a). The entropy increases quickly with the increase of the magnetic field $B$. The increase of entropy will naturally leads to large entropy force $F_e=T \partial S / \partial L$ as shown in Fig.~\ref{S12} (b). We can see that the entropic force will go to infinite when approaching the the screening distance. As discussed in Ref.\cite{Kharzeev:2014pha}, the large entropy force is considered as a important reason of driving the dissociation process. Thus, we can conclude that the production rate of quarkonium will suppress with the increase of the magnetic field $B$.

\begin{figure}[H]
    \centering
    \includegraphics[width=16cm]{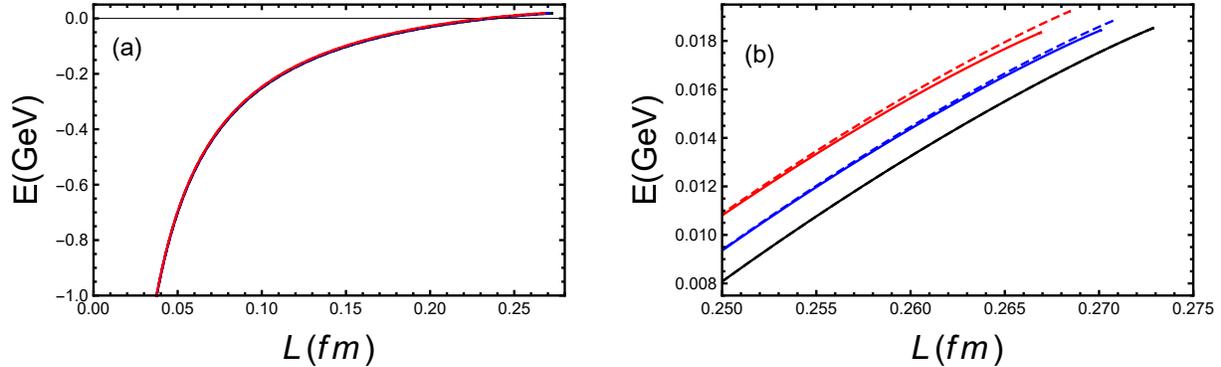}
    \caption{\label{E12} The binding energy of quarkonium on $L$ at different magnetic fields. Black line is $B = 0 \mathrm{GeV}^2$, blue line is $B = 0.1 \mathrm{GeV}^2$ and red line is $B = 0.15 \mathrm{GeV}^2$ for vanishing chemical potential. The solid line is for transverse direction and dashed line is for parallel direction. (b) is a partly enlarged view of (a).
}
\end{figure}

We compute the dependence of binding energy of heavy $Q\bar{Q}$ pair on the interquark distance $L$ at different magnetic fields $B$ in Fig.~\ref{E12}. The binding energy increases with the increase of the magnetic fields $B$. At a certain distance $L_c \leq L_S$, the free energy of the bound $Q\bar{Q}$ pair equals the free energy of an unbound $Q\bar{Q}$ pair, namely $E_{L_c} = 0$, while for larger distance the free energy of an unbound pair is smaller than that of a bound pair. However, it dose not imply the $Q\bar{Q}$ pair dissociates at this length scale. The quark-antiquark pair will become metastable even beyond $L_c$ as discussed in Ref.\cite{Ewerz:2016zsx}. When increasing $B$, the binding energy will reach to zero at smaller distance. It may indicate that the binding quarks becomes weaker at large magnetic field, especially in the transverse magnetic field.

\begin{figure}[H]
    \centering
    \includegraphics[width=16cm]{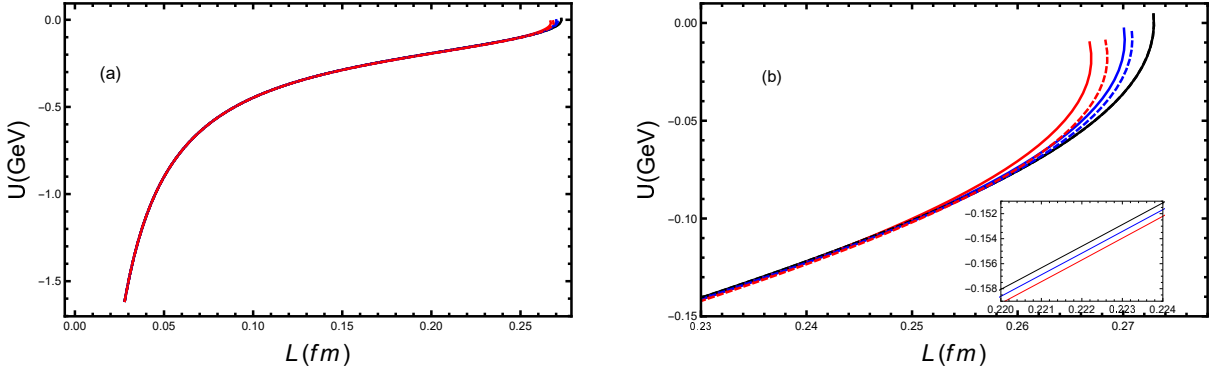}
    \caption{\label{U12}The internal energy of quarkonium on $L$ at different magnetic fields. Black line is $B = 0 \mathrm{GeV}^2$, blue line is $B = 0.1 \mathrm{GeV}^2$ and red line is $B = 0.15 \mathrm{GeV}^2$ for vanishing chemical potential. The solid line is for transverse direction and dashed line is for parallel direction. (b) is a partly enlarged view of (a). The subgraph in (b) is a partly enlarged picture of (b), we only show the magnetic field of transverse direction}.
\end{figure}

As we mentioned in the previous section, we use $U_{Q \bar{Q}}=F_{Q \bar{Q}}+T S_{Q \bar{Q}}$ to define the internal energy of $Q\bar{Q}$. The dependence of internal energy of heavy $Q\bar{Q}$  pair on the interquark distance $L$ of $Q\bar{Q}$ pair at different magnetic fields $B$ is shown in Fig.~\ref{U12}. For the small separating distance, the behavior of the internal energy is slightly suppressed. However, for the large separating distance, the internal energy will increase with the increase magnetic field. This difference may be due to the contribution of entropy at large interquark distance while $F$ is dominant at small interquark distance.

\section{Thermodynamic quantities of heavy quarkonium in EMD model with magnetic field}\label{sec:04}
In this section, we will consider the effect of dilaton field. A conformal theory often can not describe the real QCD. The dilaton introduced in the action achieved many successes in QCD phase transition, meson spectrum and other aspects. In our paper, adding a dilaton field can get the Cornell potential and correct behavior of free energy under magnetic field. Besides, a action without the dilaton can not get a proper behavior of single quark as we will discuss in the next section. Thus, as a compare, we also show the results of EMD model.

A five dimensional EMD model with Maxwell fields\cite{Bohra:2019ebj},
\begin{equation}
S= \int -\frac{1}{16\pi G_{5}}\sqrt{-g}(R-\frac{f_{1}(\phi)}{4}F_{(1)MN}F^{MN}-\frac{f_{2}(\phi)}{4}F_{(2)MN}F^{MN}-\frac{1}{2}\partial_{M}\phi\partial^{M}\phi-V(\phi) )d^{5}x.
\end{equation}
Where $F_{(1)MN}$ and $F_{(2)MN}$ are the field strength tensors, $\phi$ is the dilaton field. $f_{1}(\phi)$, $f_{2}(\phi)$ are the gauge kinetic functions. The solution of metric in string frame is
\begin{equation}
ds^{2}=\frac{L^{2}e^{2A_{s}(z)}}{z^{2}}[(-g(z)dt^{2}+\frac{1}{g(z)}dz^{2}+dy^{2}_{1}+e^{B^{2}z^{2}}(dy^{2}_{2}+dy^{2}_{3})].
\end{equation}
Magnetic field is in $y_1$ direction and $A_{s}=A(z)+\sqrt{\frac{1}{6}}\phi(z)$, with

\begin{gather}
    A(z)=-az^{2}, \\
   K_{3}=-\frac{1+\frac{\widetilde\mu^{2}}{2cL^{2}}\int_0^{z_{h}} d\xi \xi^{3} e^{-B^{2}\xi^{2}-3A(\xi)+c\xi^{2}}}{\int_0^{z_{h}} d\xi \xi^{3} e^{-B^{2}\xi^{2}-3A(\xi)}},\\
    g(z)=1+ \int_0^{z} d\xi \xi^{3} e^{-B^{2}\xi^{2}-3A(\xi)}(K_{3}+\frac{\widetilde{\mu}^{2}}{2cL^{2}e^{c\xi^{2}}}),\\
    \phi(z)=\frac{(9a-B^{2})\log(\sqrt{6a^{2}-B^{4}}\sqrt{6a^{2}z^{2}+9a^{2}-B^{4}z^{2}-B^{2}}+6a^{2}z-B^{4}z)}{\sqrt{6a^{2}-B^{4}}}
    \\+z\sqrt{6a^{2}z^{2}+9a-B^{2}(B^{2}z^{2}+1)}-\frac{(9a-B^{2})\log(\sqrt{9a-B^{2}}\sqrt{6a^{2}-B^{4}})}{\sqrt{6a^{2}-B^{4}}}.
\end{gather}

The Hawking temperature is given as
\begin{equation}
T=- \frac{z_{h}e^{A(z_{h})-B^{2}z_{h}^{2}}}{4\pi}(K_{3}+\frac{\widetilde{\mu}^{2}}{2cL^{2}}e^{c z_{h}^{2}}).
\end{equation}
In this case, the $g^{par}_1(z)$, $g^{par}_2(z)$ for parallel direction are
\begin{gather}
    g^{par}_1(z) =\frac{e^{4A_{s}}}{z^{4}}, \\
    g^{par}_2(z) =\frac{e^{4A_{s}}g(z)}{z^{4}}.
\end{gather}
And for transverse direction, $g^{tra}_1(z)$, $g^{tra}_2(z)$ are

\begin{equation}
\begin{aligned}
    g^{tra}_1(z)&=\frac{e^{4A_{s}}}{z^{4}},\\
    g^{tra}_2(z)&=\frac{e^{4A_{s}}g(z)}{z^{4}}e^{B^{2}z^{2}}.
\end{aligned}
\end{equation}

Similarly, one can use Eq.(12) to Eq.(15) in the last section to get separating distance, free energy, entropy and binding energy. Note that there is no limit of $B$ in this model, but we set the maximum of magnetic field is 0.3 $GeV^2$, which is close to magnetic field created in recent experiment.

\begin{figure} [H]
    \centering
    \includegraphics[width=16cm]{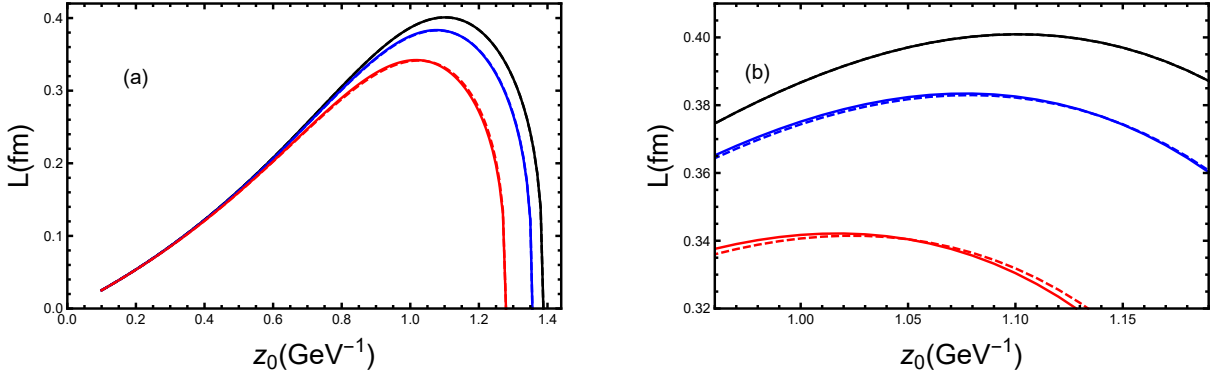}
    \caption{\label{L222}The dependence of interquark distance $L$ of $Q\bar{Q}$ pair as a function of $z_0$ for different magnetic fields. Black line is $B = 0$, blue line is $B = 0.15$ and red line is $B = 0.3$ for vanishing chemical potential. The solid line is for transverse direction and dashed line is for parallel direction. (b) is a partly enlarged view of (a).}
\end{figure}

Fig.~\ref{L222} shows the dependence of interquark distance $L$ of $Q\bar{Q}$ pair on $z_0$ at different $B=0,0.15,0.3 \mathrm{GeV}^2$. This picture also tells us that magnetic field will suppress the screening distance which seems similar to the previous case. But it is found that the magnetic field will affect the screening distance stronger with the presence of dilaton. Moreover, in the enlarged picture, we can see that the screening distance is smaller in the parallel magnetic field than that in the transverse magnetic field which is different from the EM model. Thus, we assume that dilaton will reinforce the magnetic effect and affect the spatial distribution of magnetic field.

\begin{figure}[H]
    \centering
    \includegraphics[width=16cm]{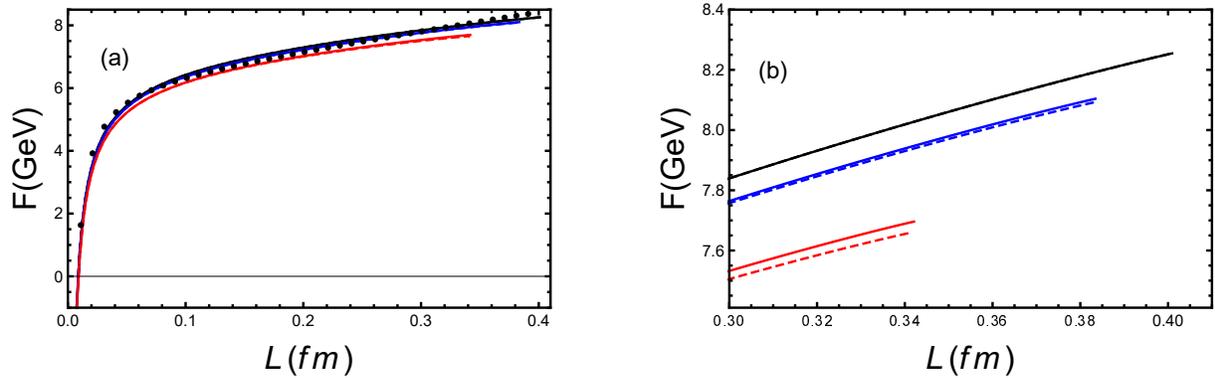}
    \caption{\label{F111}The free energy of quarkonium on $L$ for different magnetic fields. Black line is $B = 0 \mathrm{GeV}^2$, blue line is $B = 0.15 \mathrm{GeV}^2$ and red line is $B = 0.3 \mathrm{GeV}^2$ for vanishing chemical potential. The solid line is for transverse direction and dashed line is for parallel direction. The dot is a fit of the free energy with Cornell potential}. (b) is a partly enlarged view of (a).
\end{figure}

 The dependence of free energy of heavy $Q\bar{Q}$  pair on the interquark distance $L$ at different magnetic fields $B$ is plotted in Fig.~\ref{F111}. It is shown that the free energy is a Coulomb potential at small $L$ and a linear potential at large $L$ after adding the dilaton field. To be more clear, we fit the free energy at zero magnetic field with $F = 6.25 -\frac{0.515}{L} + 5.74 L$. And one can also find that the free energy is suppressed for large magnetic field $B$. Besides, we can find that the magnetic field in parallel direction will have large influence on the free energy, which is consistent with lattice results\cite{Bonati:2016kxj}.

\begin{figure}[H]
    \centering
    \includegraphics[width=16cm]{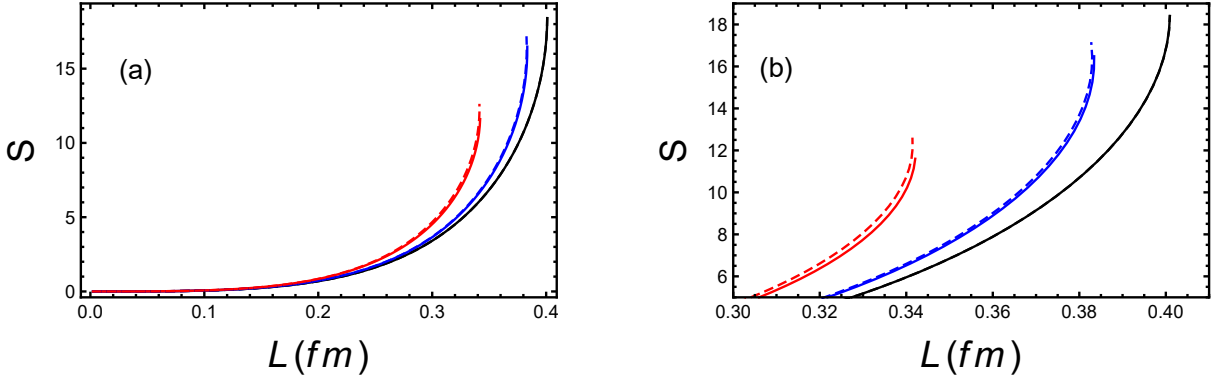}
    \caption{\label{S111}The entropy of quarkonium on $L$ at different magnetic fields. Black line is $B = 0 \mathrm{GeV}^2$, blue line is $B = 0.15 \mathrm{GeV}^2$ and red line is $B = 0.3 \mathrm{GeV}^2$ for vanishing chemical potential. The solid line is for transverse direction and dashed line is for parallel direction. (b) is a partly enlarged view of (a).}
\end{figure}

We show the dependence of entropy of heavy $Q\bar{Q}$ pair on the interquark distance $L$ at different magnetic fields $B$ in Fig.~\ref{S111}.  The entropy also increases with the increase of the magnetic field $B$ in the EMD model. This increase of entropy will lead to large entropy force. But transverse magnetic field leads to larger entropy than parallel magnetic field at large L. When approaching to screening distance $L_s$, the effect of parallel magnetic field will become large, which is different from the EM model.

\begin{figure}[H]
    \centering
    \includegraphics[width=16cm]{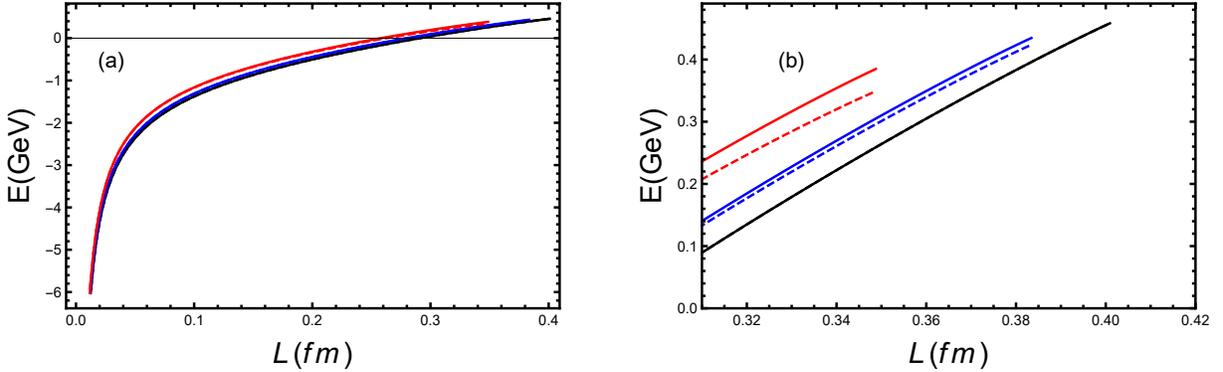}
    \caption{\label{E111}The binding energy E of quarkonium on $L$ at different magnetic fields. Black line is $B = 0 \mathrm{GeV}^2$, blue line is $B = 0.15 \mathrm{GeV}^2$ and red line is $B = 0.3 \mathrm{GeV}^2$ for vanishing chemical potential. The solid line is for transverse direction and dashed line is for parallel direction. (b) is a partly enlarged view of (a).
}
\end{figure}

We compute the dependence of binding energy of heavy $Q\bar{Q}$  pair on the interquark distance $L$ at different magnetic fields $B$ in Fig.~\ref{E111}. The binding energy increases with the increase of the magnetic field $B$. It is found that the binding energy at fixed distance becomes weaker(notice binding energy is negative) with the increase of magnetic field.

\begin{figure}[H]
    \centering
    \includegraphics[width=16cm]{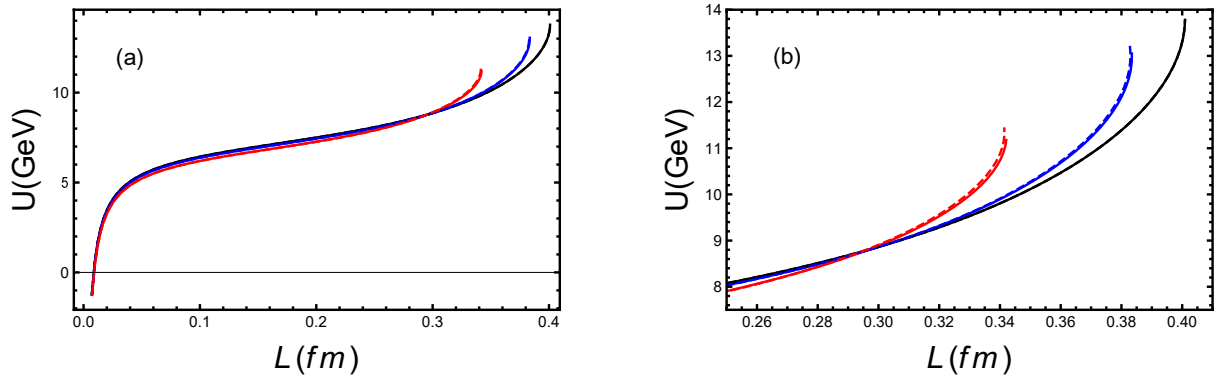}
    \caption{\label{U111}The internal energy U of quarkonium on $L$ at different magnetic fields. Black line is $B = 0 \mathrm{GeV}^2$, blue line is $B = 0.15 \mathrm{GeV}^2$ and red line is $B = 0.3 \mathrm{GeV}^2$ for vanishing chemical potential. The solid line is for transverse direction and dashed line is for parallel direction. (b) is a partly enlarged view of (a).
}
\end{figure}

The dependence of internal energy of heavy $Q\bar{Q}$ pair on the interquark distance $L$ of $Q\bar{Q}$  pair at different magnetic fields $B$ is shown in Fig.~\ref{U111}. As discussed before, for small separating distance $L$, the behavior of the internal energy is dominant by free energy. However, for large separating distance $L$, the internal energy is dominant by the entropy. Thus, the internal energy is suppressed at small $L$ and increase at large $L$ in the presence of magnetic field. And the behavior of internal energy at large $L$ is similar to entropy.

\section{Thermodynamic quantities of single quark in EMD model with magnetic field}\label{sec:05}

In the vanishing magnetic field, the free energy, entropy and internal energy of single quark can be calculated easily by considering the U-shape string approaches to the horizon $z_h$ in the conformal theory. As given in Ref.\cite{Ewerz:2016zsx}:

\begin{equation}F_{Q}=-\frac{\sqrt{\lambda}}{2} T, \quad S_{Q}=\frac{\sqrt{\lambda}}{2}, \quad U_{Q}=0\end{equation}.

Comparing with lattice results\cite{Kaczmarek:2007pb}, obviously, we find the conformal theory can't describe the single quark well. But in the EMD model, we can do further calculation. As we know, the Wilson loop is viewed as the phase factor associated to the propagation of a very massive quark in the fundamental representation of the gauge group\cite{Maldacena:1998im}. Therefore, we only emphasis the qualitative behavior with 2+1 flavor results.

\begin{figure}[H]
    \centering
    \includegraphics[width=8.5cm]{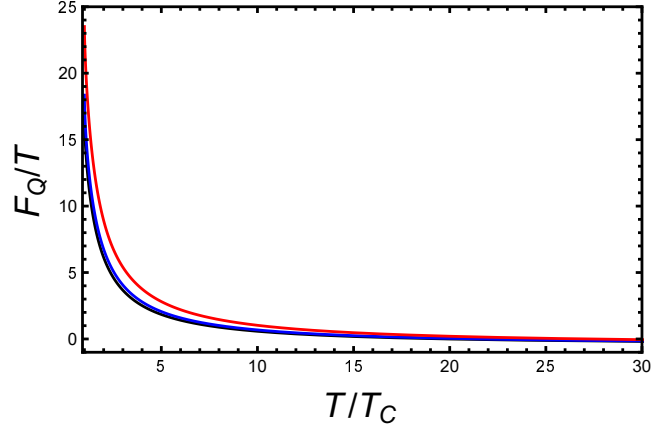}
    \caption{\label{SF111}Free energy of single quark as a function of T. Black line is $B = 0\mathrm{GeV}^2$, blue line is $B = 0.3\mathrm{GeV}^2$ and red line is $B = 0.5\mathrm{GeV}^2$ for vanishing chemical potential. }
\end{figure}

\begin{figure}[H]
    \centering
    \includegraphics[width=8.5cm]{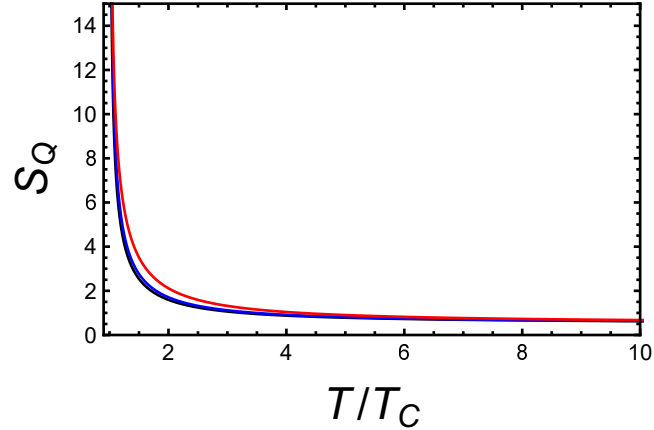}
    \caption{\label{SS111} Entropy of single quark as a function of T. Black line is $B = 0$, blue line is $B = 0.3\mathrm{GeV}^2$ and red line is $B = 0.5\mathrm{GeV}^2$ for vanishing chemical potential. }
\end{figure}

\begin{figure}[H]
    \centering
    \includegraphics[width=8.5cm]{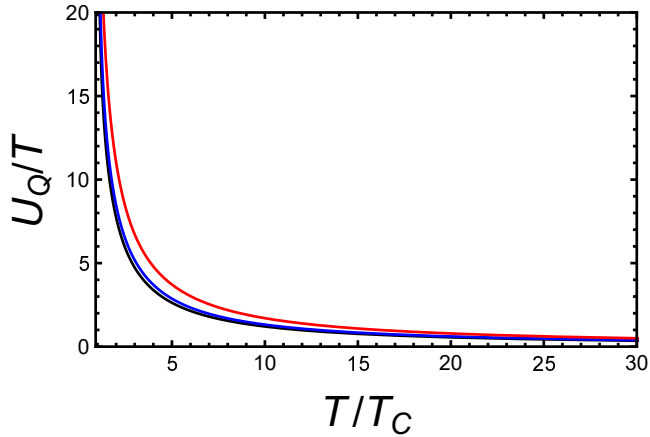}
    \caption{\label{SU111} Internal energy of single quark as a function of T. Black line is $B = 0\mathrm{GeV}^2$, blue line is $B = 0.3\mathrm{GeV}^2$ and red line is $B = 0.5\mathrm{GeV}^2$ for vanishing chemical potential. }
\end{figure}

The free energy of a single quark in this model is shown in Fig.\ref{SF111}. One can find that this figure capture the behavior of $F_{Q}/T$ in lattice QCD calculation in $B = 0$\cite{Kaczmarek:2007pb}. The increase of $B$ will lead to the increase of free energy. And the $F_{Q}/T$ will tend to the conformal case at very large $T$ limit for any given magnetic field. The single quark entropy is shown in Fig.\ref{SS111}. And one can find that the $S_{Q}$ will increase with the increase of magnetic field. At large $T$ limit, the results will tend to conformal case. The internal energy of a single quark is shown in Fig.\ref{SU111}. It is also shown that $U_Q$ will increase with the increase of the magnetic field and tend to the conformal limit at large $T$.

\section{Summary and conclusions}\label{sec:06}

In this paper, we study the free energy, entropy, binding energy, internal energy of heavy quarkonium in the 5-dimensional EM and EMD model under the magnetic field. It is found that the increase of magnetic field will suppress the screening distance, and free energy. The entropy will increase quickly with the presence of magnetic field. The induced entropic force will lead to the strong quarkonium suppression due to the entropic destruction. Then, we can see that the absolute value of binding energy(negative energy) will decrease with the increase of magnetic field, which means the binding energy of quarkonium become weaker. The internal energy is dominant by $F$ at small separating distance and dominant by $TS$ at large separating distance. Thus, we can see the internal energy will suppress at small separating distance and increase at large separating distance. Besides, the effect of magnetic field is not significant at the small separating distance.

By comparing two models, we can see the effect of dilaton on the thermodynamic qualities. The transverse and parallel magnetic field will have different effects. Thus, we assume that the dilaton will deform the space and influence the effect of magnetic field at different directions. Further, we find EMD model can realize the Cornell potential and show the magnetic field in parallel direction has large influence on potential, which is consistent with lattice results\cite{Bonati:2016kxj}. In \cite{Hasan:2017fmf}, they found the potential and binding energy is a decreasing function of magnetic field. Ref.\cite{Marasinghe:2011bt} calculates the dissociation probability increases with the magnetic field strength, which means the heavy quarkonia become unstable in the presence of magnetic field. These results also support our conclusion. Thus, in our paper, we again conform the importance of dilaton field in holographic model. Since a deformed model can describe the thermodynamic qualities of single quark at vanishing magnetic field which is qualitatively similar to the 2+1 flavor lattice QCD calculation\cite{Kaczmarek:2007pb}, we show the results of single quark and find magnetic field will enhance the free energy, entropy and internal energy in the EMD model.

\section*{Acknowledgments}
This work is partly supported by the National Science Foundation of China under Contract Nos. 11775118, 11535005.


\section*{References}
\begin{thebibliography}{99}
\bibitem{Adcox:2004mh}
  K.~Adcox {\it et al.} [PHENIX Collaboration],
  %``Formation of dense partonic matter in relativistic nucleus-nucleus collisions at RHIC: Experimental evaluation by the PHENIX collaboration,''
  Nucl.\ Phys.\ A {\bf 757}, 184 (2005)
%\cite{Adams:2005dq}
\bibitem{Adams:2005dq}
  J.~Adams {\it et al.} [STAR Collaboration],
  %``Experimental and theoretical challenges in the search for the quark gluon plasma: The STAR Collaboration's critical assessment of the evidence from RHIC collisions,''
  Nucl.\ Phys.\ A {\bf 757}, 102 (2005)
%\cite{Arsene:2004fa}
\bibitem{Arsene:2004fa}
  I.~Arsene {\it et al.} [BRAHMS Collaboration],
  %``Quark gluon plasma and color glass condensate at RHIC? The Perspective from the BRAHMS experiment,''
  Nucl.\ Phys.\ A {\bf 757}, 1 (2005)
%\cite{Miller:2007ri}
\bibitem{Miller:2007ri}
  M.~L.~Miller, K.~Reygers, S.~J.~Sanders and P.~Steinberg,
  %``Glauber modeling in high energy nuclear collisions,''
  Ann.\ Rev.\ Nucl.\ Part.\ Sci.\  {\bf 57}, 205 (2007)
  doi:10.1146/annurev.nucl.57.090506.123020
%\cite{Zollner:2020nnt}
\bibitem{Zollner:2020nnt}
R.~Z\"ollner and B.~K\"ampfer,
%``Quarkonia formation in a holographic gravity-dilaton background describing QCD thermodynamics,''
[arXiv:2007.14287 [hep-ph]].
%0 citations counted in INSPIRE as of 26 Sep 2020
%\cite{Matsui:1986dk}
\bibitem{Matsui:1986dk}
T.~Matsui and H.~Satz,
%``$J/\psi$ Suppression by Quark-Gluon Plasma Formation,''
Phys. Lett. B \textbf{178}, 416-422 (1986)
doi:10.1016/0370-2693(86)91404-8
%3036 citations counted in INSPIRE as of 02 Jun 2020

%\cite{Maldacena:1997re}
\bibitem{Maldacena:1997re}
  J.~M.~Maldacena,
  %``The Large N limit of superconformal field theories and supergravity,''
  Int.\ J.\ Theor.\ Phys.\  {\bf 38}, 1113 (1999)
  [Adv.\ Theor.\ Math.\ Phys.\  {\bf 2}, 231 (1998)]
  doi:10.1023/A:1026654312961, 10.4310/ATMP.1998.v2.n2.a1
  [hep-th/9711200].
%\cite{Witten:1998qj}
\bibitem{Witten:1998qj}
  E.~Witten,
  %``Anti-de Sitter space and holography,''
  Adv.\ Theor.\ Math.\ Phys.\  {\bf 2}, 253 (1998)
  doi:10.4310/ATMP.1998.v2.n2.a2
  [hep-th/9802150]


%\cite{Gubser:1998bc}
\bibitem{Gubser:1998bc}
S.~Gubser, I.~R.~Klebanov and A.~M.~Polyakov,
%``Gauge theory correlators from noncritical string theory,''
Phys. Lett. B \textbf{428}, 105-114 (1998)
doi:10.1016/S0370-2693(98)00377-3
[arXiv:hep-th/9802109 [hep-th]].
%8606 citations counted in INSPIRE as of 02 Jun 2020

%\cite{Kharzeev:2007jp}
\bibitem{Kharzeev:2007jp}
D.~E.~Kharzeev, L.~D.~McLerran and H.~J.~Warringa,
%``The Effects of topological charge change in heavy ion collisions: 'Event by event P and CP violation',''
Nucl. Phys. A \textbf{803}, 227-253 (2008)
doi:10.1016/j.nuclphysa.2008.02.298
[arXiv:0711.0950 [hep-ph]].
%1330 citations counted in INSPIRE as of 03 Jun 2020

%\cite{Kharzeev:2004ey}
\bibitem{Kharzeev:2004ey}
D.~Kharzeev,
%``Parity violation in hot QCD: Why it can happen, and how to look for it,''
Phys. Lett. B \textbf{633}, 260-264 (2006)
doi:10.1016/j.physletb.2005.11.075
[arXiv:hep-ph/0406125 [hep-ph]].
%472 citations counted in INSPIRE as of 03 Jun 2020


%\cite{Fukushima:2008xe}
\bibitem{Fukushima:2008xe}
K.~Fukushima, D.~E.~Kharzeev and H.~J.~Warringa,
%``The Chiral Magnetic Effect,''
Phys. Rev. D \textbf{78}, 074033 (2008)
doi:10.1103/PhysRevD.78.074033
[arXiv:0808.3382 [hep-ph]].
%1343 citations counted in INSPIRE as of 03 Jun 2020

%\cite{Gatto:2012sp}
\bibitem{Gatto:2012sp}
R.~Gatto and M.~Ruggieri,
%``Quark Matter in a Strong Magnetic Background,''
Lect. Notes Phys. \textbf{871}, 87-119 (2013)
%doi:10.1007/978-3-642-37305-3_4
[arXiv:1207.3190 [hep-ph]].
%65 citations counted in INSPIRE as of 03 Jun 2020

%\cite{Abelev:2009ac}
\bibitem{Abelev:2009ac}
B.~Abelev \textit{et al.} [STAR],
%``Azimuthal Charged-Particle Correlations and Possible Local Strong Parity Violation,''
Phys. Rev. Lett. \textbf{103}, 251601 (2009)
doi:10.1103/PhysRevLett.103.251601
[arXiv:0909.1739 [nucl-ex]].
%464 citations counted in INSPIRE as of 03 Jun 2020

%\cite{Abelev:2009ad}
\bibitem{Abelev:2009ad}
B.~Abelev \textit{et al.} [STAR],
%``Observation of charge-dependent azimuthal correlations and possible local strong parity violation in heavy ion collisions,''
Phys. Rev. C \textbf{81}, 054908 (2010)
doi:10.1103/PhysRevC.81.054908
[arXiv:0909.1717 [nucl-ex]].
%343 citations counted in INSPIRE as of 03 Jun 2020

%\cite{Adamczyk:2013hsi}
\bibitem{Adamczyk:2013hsi}
L.~Adamczyk \textit{et al.} [STAR],
%``Fluctuations of charge separation  perpendicular to the event plane and local parity violation in $\sqrt{s_{NN}}=200$ GeV Au+Au  collisions at the BNL Relativistic Heavy Ion Collider,''
Phys. Rev. C \textbf{88}, no.6, 064911 (2013)
doi:10.1103/PhysRevC.88.064911
[arXiv:1302.3802 [nucl-ex]].
%87 citations counted in INSPIRE as of 03 Jun 2020

%\cite{Adamczyk:2013kcb}
\bibitem{Adamczyk:2013kcb}
L.~Adamczyk \textit{et al.} [STAR],
%``Measurement of charge multiplicity asymmetry correlations in high-energy nucleus-nucleus collisions at $\sqrt{{s}_{NN}} =$ 200 GeV,''
Phys. Rev. C \textbf{89}, no.4, 044908 (2014)
doi:10.1103/PhysRevC.89.044908
[arXiv:1303.0901 [nucl-ex]].
%64 citations counted in INSPIRE as of 03 Jun 2020

%\cite{Hirano:2010je}
\bibitem{Hirano:2010je}
T.~Hirano, P.~Huovinen and Y.~Nara,
%``Elliptic flow in Pb+Pb collisions at $\sqrt{s_{NN}} = 2.76$~TeV: hybrid model assessment of the first data,''
Phys. Rev. C \textbf{84}, 011901(R) (2011)
doi:10.1103/PhysRevC.84.011901(R)
[arXiv:1012.3955 [nucl-th]].
%61 citations counted in INSPIRE as of 03 Jun 2020

%\cite{Abelev:2012pa}
\bibitem{Abelev:2012pa}
B.~Abelev \textit{et al.} [ALICE],
%``Charge separation relative to the reaction plane in Pb-Pb collisions at $\sqrt{s_{NN}}= 2.76$ TeV,''
Phys. Rev. Lett. \textbf{110}, no.1, 012301 (2013)
doi:10.1103/PhysRevLett.110.012301
[arXiv:1207.0900 [nucl-ex]].
%246 citations counted in INSPIRE as of 03 Jun 2020

%\cite{Mo:2013qya}
\bibitem{Mo:2013qya}
Y.~J.~Mo, S.~Q.~Feng and Y.~F.~Shi,
%``Effect of the Wood-Saxon nucleon distribution on the chiral magnetic field in relativistic heavy-ion collisions,''
Phys. Rev. C \textbf{88}, no.2, 024901 (2013)
doi:10.1103/PhysRevC.88.024901
[arXiv:1308.4289 [hep-ph]].
%16 citations counted in INSPIRE as of 03 Jun 2020

%\cite{Zhong:2014cda}
\bibitem{Zhong:2014cda}
Y.~Zhong, C.~B.~Yang, X.~Cai and S.~Q.~Feng,
%``A systematic study of magnetic field in Relativistic Heavy-ion Collisions in the RHIC and LHC energy regions,''
Adv. High Energy Phys. \textbf{2014}, 193039 (2014)
doi:10.1155/2014/193039
[arXiv:1408.5694 [hep-ph]].
%27 citations counted in INSPIRE as of 03 Jun 2020

%\cite{Zhong:2014sua}
\bibitem{Zhong:2014sua}
Y.~Zhong, C.~B.~Yang, X.~Cai and S.~Q.~Feng,
%``Spatial distributions of magnetic field in the RHIC and LHC energy regions,''
Chin. Phys. C \textbf{39}, no.10, 104105 (2015)
doi:10.1088/1674-1137/39/10/104105
[arXiv:1410.6349 [hep-ph]].
%11 citations counted in INSPIRE as of 03 Jun 2020

%\cite{Feng:2016srp}
\bibitem{Feng:2016srp}
S.~Q.~Feng, L.~Pei, F.~Sun, Y.~Zhong and Z.~B.~Yin,
%``Estimation of the chiral magnetic effect considering the magnetic field response of the QGP medium,''
Chin. Phys. C \textbf{42}, no.5, 054102 (2018)
doi:10.1088/1674-1137/42/5/054102
[arXiv:1609.07550 [nucl-th]].
%1 citations counted in INSPIRE as of 03 Jun 2020


%\cite{Dudal:2015wfn}
\bibitem{Dudal:2015wfn}
D.~Dudal, D.~R.~Granado and T.~G.~Mertens,
%``No inverse magnetic catalysis in the QCD hard and soft wall models,''
Phys. Rev. D \textbf{93}, no.12, 125004 (2016)
doi:10.1103/PhysRevD.93.125004
[arXiv:1511.04042 [hep-th]].
%46 citations counted in INSPIRE as of 03 Jun 2020

%\cite{Mamo:2015dea}
\bibitem{Mamo:2015dea}
K.~A.~Mamo,
%``Inverse magnetic catalysis in holographic models of QCD,''
JHEP \textbf{05}, 121 (2015)
doi:10.1007/JHEP05(2015)121
[arXiv:1501.03262 [hep-th]].
%56 citations counted in INSPIRE as of 03 Jun 2020


%\cite{Rougemont:2015oea}
\bibitem{Rougemont:2015oea}
R.~Rougemont, R.~Critelli and J.~Noronha,
%``Holographic calculation of the QCD crossover temperature in a magnetic field,''
Phys. Rev. D \textbf{93}, no.4, 045013 (2016)
doi:10.1103/PhysRevD.93.045013
[arXiv:1505.07894 [hep-th]].
%61 citations counted in INSPIRE as of 03 Jun 2020





%\cite{Li:2016gfn}
\bibitem{Li:2016gfn}
  D.~Li, M.~Huang, Y.~Yang and P.~H.~Yuan,
  %``Inverse Magnetic Catalysis in the Soft-Wall Model of AdS/QCD,''
  JHEP {\bf 1702}, 030 (2017)
  doi:10.1007/JHEP02(2017)030
  [arXiv:1610.04618 [hep-th]].

%\cite{Rodrigues:2017iqi}
\bibitem{Rodrigues:2017iqi}
D.~M.~Rodrigues, E.F.~Capossoli and H.~Boschi-Filho,
%``Magnetic catalysis and inverse magnetic catalysis in ( 2+1 )-dimensional gauge theories from holographic models,''
Phys. Rev. D \textbf{97}, no.12, 126001 (2018)
doi:10.1103/PhysRevD.97.126001
[arXiv:1710.07310 [hep-th]].
%10 citations counted in INSPIRE as of 03 Jun 2020

%\cite{Rodrigues:2018pep}
\bibitem{Rodrigues:2018pep}
D.~M.~Rodrigues, D.~Li, E.F.~Capossoli and H.~Boschi-Filho,
%``Chiral symmetry breaking and restoration in 2+1 dimensions from holography: Magnetic and inverse magnetic catalysis,''
Phys. Rev. D \textbf{98}, no.10, 106007 (2018)
doi:10.1103/PhysRevD.98.106007
[arXiv:1807.11822 [hep-th]].
%8 citations counted in INSPIRE as of 03 Jun 2020

%\cite{He:2020fdi}
\bibitem{He:2020fdi}
S.~He, Y.~Yang and P.~H.~Yuan,
%``Analytic Study of Magnetic Catalysis in Holographic QCD,''
[arXiv:2004.01965 [hep-th]].
%1 citations counted in INSPIRE as of 03 Jun 2020


%\cite{Dudal:2016joz}
\bibitem{Dudal:2016joz}
D.~Dudal and S.~Mahapatra,
%``Confining gauge theories and holographic entanglement entropy with a magnetic field,''
JHEP \textbf{04}, 031 (2017)
doi:10.1007/JHEP04(2017)031
[arXiv:1612.06248 [hep-th]].
%26 citations counted in INSPIRE as of 03 Jun 2020

%\cite{Fujita:2020qvp}
\bibitem{Fujita:2020qvp}
M.~Fujita, S.~He and Y.~Sun,
%``Thermodynamical property of entanglement entropy and deconfinement phase transition,''
[arXiv:2005.01048 [hep-th]].
%0 citations counted in INSPIRE as of 03 Jun 2020

%\cite{Zhu:2019ujc}
\bibitem{Zhu:2019ujc}
Z.~R.~Zhu, S.~Q.~Feng, Y.~F.~Shi and Y.~Zhong,
%``Energy loss of heavy and light quarks in holographic magnetized background,''
Phys. Rev. D \textbf{99}, no.12, 126001 (2019)
doi:10.1103/PhysRevD.99.126001
[arXiv:1901.09304 [hep-ph]].
%5 citations counted in INSPIRE as of 03 Jun 2020




%\cite{Dudal:2018rki}
\bibitem{Dudal:2018rki}
  D.~Dudal and T.~G.~Mertens,
  %``Holographic estimate of heavy quark diffusion in a magnetic field,''
  Phys.\ Rev.\ D {\bf 97}, no. 5, 054035 (2018)
  doi:10.1103/PhysRevD.97.054035
  [arXiv:1802.02805 [hep-th]].
%\cite{Braga:2018zlu}
\bibitem{Braga:2018zlu}
  N.~R.~F.~Braga and L.~F.~Ferreira,
  %``Heavy meson dissociation in a plasma with magnetic fields,''
  Phys.\ Lett.\ B {\bf 783}, 186 (2018)
  doi:10.1016/j.physletb.2018.06.053
%\cite{DElia:2018xwo}
\bibitem{DElia:2018xwo}
  M.~D'Elia, F.~Manigrasso, F.~Negro and F.~Sanfilippo,
  %``QCD phase diagram in a magnetic background for different values of the pion mass,''
  Phys.\ Rev.\ D {\bf 98}, no. 5, 054509 (2018)
  doi:10.1103/PhysRevD.98.054509
  [arXiv:1808.07008 [hep-lat]].

%%%%%%%%%%%%%%%%%%%%%%%%%%%%%%%%%%%%%%%%%%%%%%%%%%%%%%%%%%%%%%%%%%xinjiaru
%\cite{Fukushima:2012kc}
\bibitem{Fukushima:2012kc}
K.~Fukushima and Y.~Hidaka,
%``Magnetic Catalysis Versus Magnetic Inhibition,''
Phys. Rev. Lett. \textbf{110}, no.3, 031601 (2013)
doi:10.1103/PhysRevLett.110.031601
[arXiv:1209.1319 [hep-ph]]
%\cite{Ferreira:2014kpa}
\bibitem{Ferreira:2014kpa}
  M.~Ferreira, P.~Costa, O.~Lourenco, T.~Frederico and C.~Providencia,
  %``Inverse magnetic catalysis in the (2+1)-flavor Nambu-Jona-Lasinio and Polyakov-Nambu-Jona-Lasinio models,''
  Phys.\ Rev.\ D {\bf 89} (2014) no.11,  116011
  doi:10.1103/PhysRevD.89.116011
  [arXiv:1404.5577 [hep-ph]].
%\cite{Bali:2011qj}
\bibitem{Bali:2011qj}
G.~Bali, F.~Bruckmann, G.~Endrodi, Z.~Fodor, S.~Katz, S.~Krieg, A.~Schafer and K.~Szabo,
%``The QCD phase diagram for external magnetic fields,''
JHEP \textbf{02}, 044 (2012)
doi:10.1007/JHEP02(2012)044
[arXiv:1111.4956 [hep-lat]].
%\cite{Evans:2016jzo}
\bibitem{Evans:2016jzo}
  N.~Evans, C.~Miller and M.~Scott,
  %``Inverse Magnetic Catalysis in Bottom-Up Holographic QCD,''
  Phys.\ Rev.\ D {\bf 94}, no. 7, 074034 (2016)
  doi:10.1103/PhysRevD.94.074034
  [arXiv:1604.06307 [hep-ph]].
  %%CITATION = doi:10.1103/PhysRevD.94.074034;%%
%\cite{Ballon-Bayona:2017dvv}
\bibitem{Ballon-Bayona:2017dvv}
  A.~Ballon-Bayona, M.~Ihl, J.~P.~Shock and D.~Zoakos,
  %``A universal order parameter for Inverse Magnetic Catalysis,''
  JHEP {\bf 1710}, 038 (2017)
  doi:10.1007/JHEP10(2017)038
  [arXiv:1706.05977 [hep-th]].

%\cite{Gursoy:2016ofp}
\bibitem{Gursoy:2016ofp}
  U.~Gürsoy, I.~Iatrakis, M.~Järvinen and G.~Nijs,
  %``Inverse Magnetic Catalysis from improved Holographic QCD in the Veneziano limit,''
  JHEP {\bf 1703}, 053 (2017)
  doi:10.1007/JHEP03(2017)053
  [arXiv:1611.06339 [hep-th]].
  %%CITATION = doi:10.1007/JHEP03(2017)053;%%
%\cite{Gursoy:2017wzz}
\bibitem{Gursoy:2017wzz}
  U.~Gursoy, M.~Jarvinen and G.~Nijs,
  %``Holographic QCD in the Veneziano Limit at a Finite Magnetic Field and Chemical Potential,''
  Phys.\ Rev.\ Lett.\  {\bf 120}, no. 24, 242002 (2018)
  doi:10.1103/PhysRevLett.120.242002
  [arXiv:1707.00872 [hep-th]].
  %%CITATION = doi:10.1103/PhysRevLett.120.242002;%%
%\cite{Dudal:2018rki}



\bibitem{Andreev:2006eh}
  O.~Andreev and V.~I.~Zakharov,
  %``The Spatial String Tension, Thermal Phase Transition, and AdS/QCD,''
  Phys.\ Lett.\ B {\bf 645}, 437 (2007)
  doi:10.1016/j.physletb.2007.01.002
  [hep-ph/0607026].


%\cite{Colangelo:2010pe}
\bibitem{Colangelo:2010pe}
  P.~Colangelo, F.~Giannuzzi and S.~Nicotri,
  %``Holography, Heavy-Quark Free Energy, and the QCD Phase Diagram,''
  Phys.\ Rev.\ D {\bf 83}, 035015 (2011)
  doi:10.1103/PhysRevD.83.035015
  [arXiv:1008.3116 [hep-ph]].

\bibitem{Li:2011hp}
  D.~Li, S.~He, M.~Huang and Q.~S.~Yan,
  %``Thermodynamics of deformed AdS$_5$ model with a positive/negative quadratic correction in graviton-dilaton system,''
  JHEP {\bf 1109}, 041 (2011)
  doi:10.1007/JHEP09(2011)041
  [arXiv:1103.5389 [hep-th]].
  %%CITATION = doi:10.1007/JHEP09(2011)041;%%
%\cite{Dudal:2016joz}

%\cite{Herzog:2006ra}
\bibitem{Herzog:2006ra}
  C.~P.~Herzog,
  %``A Holographic Prediction of the Deconfinement Temperature,''
  Phys.\ Rev.\ Lett.\  {\bf 98}, 091601 (2007)
  doi:10.1103/PhysRevLett.98.091601
  [hep-th/0608151].

%\cite{Ewerz:2016zsx}
\bibitem{Ewerz:2016zsx}
  C.~Ewerz, O.~Kaczmarek and A.~Samberg,
  %``Free Energy of a Heavy Quark-Antiquark Pair in a Thermal Medium from AdS/CFT,''
  JHEP {\bf 1803}, 088 (2018)
  doi:10.1007/JHEP03(2018)088
  [arXiv:1605.07181 [hep-th]].

%\cite{Kaczmarek:2002mc}
\bibitem{Kaczmarek:2002mc}
  O.~Kaczmarek, F.~Karsch, P.~Petreczky and F.~Zantow,
  %``Heavy quark anti-quark free energy and the renormalized Polyakov loop,''
  Phys.\ Lett.\ B {\bf 543}, 41 (2002)
  doi:10.1016/S0370-2693(02)02415-2
  [hep-lat/0207002].

%\cite{Kaczmarek:2004gv}
\bibitem{Kaczmarek:2004gv}
  O.~Kaczmarek, F.~Karsch, F.~Zantow and P.~Petreczky,
  %``Static quark anti-quark free energy and the running coupling at finite temperature,''
  Phys.\ Rev.\ D {\bf 70}, 074505 (2004)

%\cite{Kharzeev:2014pha}
\bibitem{Kharzeev:2014pha}
D.~E.~Kharzeev,
%``Deconfinement as an entropic self-destruction: a solution for the quarkonium suppression puzzle?,''
Phys.\ Rev.\ D \textbf{90} (2014) no.7, 074007
doi:10.1103/PhysRevD.90.074007
[arXiv:1409.2496 [hep-ph]].

%\cite{Satz:2015jsa}
\bibitem{Satz:2015jsa}
H.~Satz,
%``Quarkonium Binding and Entropic Force,''
Eur. Phys. J. C \textbf{75}, no.5, 193 (2015)
doi:10.1140/epjc/s10052-015-3424-7
[arXiv:1501.03940 [hep-ph]].
%12 citations counted in INSPIRE as of 03 Jun 2020


  %\cite{Critelli:2016cvq}
\bibitem{Critelli:2016cvq}
R.~Critelli, R.~Rougemont, S.~I.~Finazzo and J.~Noronha,
%``Polyakov loop and heavy quark entropy in strong magnetic fields from holographic black hole engineering,''
Phys. Rev. D \textbf{94}, no.12, 125019 (2016)
doi:10.1103/PhysRevD.94.125019
[arXiv:1606.09484 [hep-ph]].


%\cite{Dudal:2017max}
\bibitem{Dudal:2017max}
D.~Dudal and S.~Mahapatra,
%``Thermal entropy of a quark-antiquark pair above and below deconfinement from a dynamical holographic QCD model,''
Phys. Rev. D \textbf{96}, no.12, 126010 (2017)
doi:10.1103/PhysRevD.96.126010
[arXiv:1708.06995 [hep-th]].
%12 citations counted in INSPIRE as of 03 Jun 2020


\bibitem{Petreczky:2005bd}
  P.~Petreczky,
  %``Heavy quark potentials and quarkonia binding,''
  Eur.\ Phys.\ J.\ C {\bf 43}, 51 (2005)
  doi:10.1140/epjc/s2005-02261-6
  [hep-lat/0502008].
\bibitem{Fadafan:2015ynz}
  K.B.~Fadafan and S.~K.~Tabatabaei,
  %``Entropic destruction of a moving heavy quarkonium,''
  Phys.\ Rev.\ D {\bf 94}, no. 2, 026007 (2016)
  doi:10.1103/PhysRevD.94.026007
  [arXiv:1512.08254 [hep-ph]].

\bibitem{Chen:2017lsf}
  X.~Chen, S.~Q.~Feng, Y.~F.~Shi and Y.~Zhong,
  %``Moving heavy quarkonium entropy, effective string tension, and the QCD phase diagram,''
  Phys.\ Rev.\ D {\bf 97}, no. 6, 066015 (2018)
  doi:10.1103/PhysRevD.97.066015
  [arXiv:1710.00465 [hep-ph]].

\bibitem{Zhang:2016fwr}
Z.~Zhang, C.~Ma, D.~Hou and G.~Chen,
%``Entropic destruction of a rotating heavy quarkonium,''
[arXiv:1611.08011 [hep-th]].
%4 citations counted in INSPIRE as of 11 Apr 2020


%\cite{Zhang:2017izi}
\bibitem{Zhang:2017izi}
Z.~Zhang, D.~Hou, Z.~Luo and G.~Chen,
%``Entropic Destruction of Heavy Quarkonium from a Deformed AdS5 Model,''
Adv.\ High Energy Phys.\  \textbf{2017} (2017), 8910210
doi:10.1155/2017/8910210
[arXiv:1701.06147 [hep-ph]].
%2 citations counted in INSPIRE as of 11 Apr 2020


%\cite{Zhang:2020upv}
\bibitem{Zhang:2020upv}
Z.~Zhang and D.~Hou,
%``Entropic destruction of heavy quarkonium in quark-gluon plasma with gluon condensate,''
Phys.\ Lett.\ B \textbf{803} (2020), 135301
doi:10.1016/j.physletb.2020.135301
%0 citations counted in INSPIRE as of 11 Apr 2020

%\cite{Arefeva:2018hyo}
\bibitem{Arefeva:2018hyo}
I.~Aref'eva and K.~Rannu,
%``Holographic Anisotropic Background with Confinement-Deconfinement Phase Transition,''
JHEP \textbf{05}, 206 (2018)
doi:10.1007/JHEP05(2018)206
[arXiv:1802.05652 [hep-th]].
%25 citations counted in INSPIRE as of 12 Aug 2020

%\cite{Arefeva:2018cli}
\bibitem{Arefeva:2018cli}
I.~Aref'eva, K.~Rannu and P.~Slepov,
%``Orientation Dependence of Confinement-Deconfinement Phase Transition in Anisotropic Media,''
Phys. Lett. B \textbf{792}, 470-475 (2019)
doi:10.1016/j.physletb.2019.04.012
[arXiv:1808.05596 [hep-th]].
%13 citations counted in INSPIRE as of 12 Aug 2020




%\cite{Feng:2019boe}
\bibitem{Feng:2019boe}
S.~Q.~Feng, Y.~Q.~Zhao and X.~Chen,
%``Systematical study of thermal width of heavy quarkonia in a finite temperature magnetized background from holography,''
Phys. Rev. D \textbf{101}, no.2, 026023 (2020)
doi:10.1103/PhysRevD.101.026023
[arXiv:1910.05668 [hep-ph]].
%0 citations counted in INSPIRE as of 03 Jun 2020

%\cite{Zhu:2019igg}
\bibitem{Zhu:2019igg}
Z.~R.~Zhu, D.~f.~Hou and X.~Chen,
%``Potential analysis of holographic Schwinger effect in the magnetized background,''
[arXiv:1912.05806 [hep-ph]].
%1 citations counted in INSPIRE as of 03 Jun 2020


%\cite{Bali:2014kia}
\bibitem{Bali:2014kia}
G.~Bali, F.~Bruckmann, G.~Endrödi, S.~Katz and A.~Schäfer,
%``The QCD equation of state in background magnetic fields,''
JHEP \textbf{08}, 177 (2014)
doi:10.1007/JHEP08(2014)177
[arXiv:1406.0269 [hep-lat]].
%146 citations counted in INSPIRE as of 07 Jun 2020

%\cite{Bonati:2016kxj}
\bibitem{Bonati:2016kxj}
C.~Bonati, M.~D'Elia, M.~Mariti, M.~Mesiti, F.~Negro, A.~Rucci and F.~Sanfilippo,
%``Magnetic field effects on the static quark potential at zero and finite temperature,''
Phys. Rev. D \textbf{94}, no.9, 094007 (2016)
doi:10.1103/PhysRevD.94.094007
[arXiv:1607.08160 [hep-lat]].
%39 citations counted in INSPIRE as of 10 Aug 2020


%\cite{Bohra:2019ebj}
\bibitem{Bohra:2019ebj}
H.~Bohra, D.~Dudal, A.~Hajilou and S.~Mahapatra,
%``Anisotropic string tensions and inversely magnetic catalyzed deconfinement from a dynamical AdS/QCD model,''
Phys. Lett. B \textbf{801}, 135184 (2020)
doi:10.1016/j.physletb.2019.135184
[arXiv:1907.01852 [hep-th]].
%7 citations counted in INSPIRE as of 03 Jun 2020

%\cite{Kaczmarek:2007pb}
\bibitem{Kaczmarek:2007pb}
O.~Kaczmarek,
%``Screening at finite temperature and density,''
PoS \textbf{CPOD07}, 043 (2007)
doi:10.22323/1.047.0043
[arXiv:0710.0498 [hep-lat]].
%48 citations counted in INSPIRE as of 03 Jun 2020

%\cite{Maldacena:1998im}
\bibitem{Maldacena:1998im}
J.~M.~Maldacena,
%``Wilson loops in large N field theories,''
Phys. Rev. Lett. \textbf{80}, 4859-4862 (1998)
doi:10.1103/PhysRevLett.80.4859
[arXiv:hep-th/9803002 [hep-th]].
%1694 citations counted in INSPIRE as of 12 Aug 2020

%\cite{Hasan:2017fmf}
\bibitem{Hasan:2017fmf}
M.~Hasan, B.~Chatterjee and B.~K.~Patra,
%``Heavy Quark Potential in a static and strong homogeneous magnetic field,''
Eur. Phys. J. C \textbf{77} (2017) no.11, 767
doi:10.1140/epjc/s10052-017-5346-z
[arXiv:1703.10508 [hep-ph]].
%29 citations counted in INSPIRE as of 26 Sep 2020

%\cite{Marasinghe:2011bt}
\bibitem{Marasinghe:2011bt}
K.~Marasinghe and K.~Tuchin,
%``Quarkonium dissociation in quark-gluon plasma via ionization in magnetic field,''
Phys. Rev. C \textbf{84} (2011), 044908
doi:10.1103/PhysRevC.84.044908
[arXiv:1103.1329 [hep-ph]].
%51 citations counted in INSPIRE as of 26 Sep 2020


\end{thebibliography}
\end{document}